\documentclass[twocolumn,preprintnumbers]{revtex4}
\usepackage{amssymb}

\usepackage{amsmath,amsfonts,amssymb,epsfig,color}

\begin{document}

\title{Transition probabilities in the $U(6)$ limit of the \\ Symplectic Interacting Vector Boson Model}
\author{H. G. Ganev and A. I. Georgieva}
\affiliation{Institute of
Nuclear Research and Nuclear Energy, Bulgarian Academy of
Sciences, Sofia 1784, Bulgaria}

%\date{\today}

\setcounter{MaxMatrixCols}{10}

\begin{abstract}
The tensor properties of the algebra generators and the basis
are determined in respect to the reduction chain $Sp(12,R) \supset U(6)%
\supset U(3)\otimes U(2)\supset O(3)\otimes (U(1)\otimes U(1))$,
which defines one of the dynamical symmetries  of the Interacting
Vector Boson Model. The action of the $Sp(12,R)$ generators as
transition operators between the basis states is presented.
Analytical expressions for their matrix elements in the
symmetry-adapted basis are obtained. As an example the matrix
elements of the $E2$ transition operator between collective states
of the ground band are determined and compared  with the
experimental data for the corresponding intraband transition
probabilities of nuclei in the actinide and rare earth region. On
the basis of this application the important role of the symplectic
extension of the model is analyzed.
\end{abstract}
\maketitle
 PACS {23.20.-g, 23.20.Js, 21.60.Fw, 21.10.Re, 27.70.+q, 27.80.+w}
%\end{start}

\section[]{Introduction\label{s1}}

In the algebraic models the use of the dynamical symmetries defined
by a certain reduction chain of the group of dynamical symmetry,
yields exact solutions for the eigenvalues and eigenfunctions of the
model Hamiltonian, which is constructed from the invariant operators
of the subgroups in the chain.

Something more, it is very simple and straightforward to calculate
matrix elements of transition operators between the eigenstates,
as both - the basis states and the operators, can be defined as
tensor operators in respect to the considered dynamical symmetry.
Then the calculation of matrix elements is simplified by the use
of a respective generalization of the Wigner-Eckart theorem, which
requires the calculation of the isoscalar factors and reduced
matrix elements. By definition such matrix elements give the
transition probabilities between the collective states attributed
to the basis states of the Hamiltonian.

The comparison of the experimental data with the calculated
transition probabilities is one of the best tests of the validity
of the considered algebraic model. With the aim of such
applications of the symplectic extension of one of the dynamical
symmetries in IVBM, we develop in this paper a practical
mathematical approach for explicit evaluation of the matrix
elements of transitional operators in the model.

The algebraic Interacting Vector Bosons Model (IVBM) was developed
\cite{IVBM} initially for the description of the low lying bands of
the well deformed even-even nuclei \cite{IVBMrl}. Recently this
approach was adapted to incorporate the newly observed higher
collective states, both in the first positive and negative parity
bands \cite{Sp12U6} by considering the basis states as
\textquotedblright yrast\textquotedblright\ states for the different
values of the number of bosons $N,$ that built them. This was
achieved by extending the dynamical symmetry group $U(6)$ to the noncompact $%
Sp(12,R).$ The excellent results obtained for the energy spectrum
motivated the present investigation of the transition probabilities
in the framework of the generalized IVBM with $Sp(12,R)$ as a group
of dynamical symmetry. Thus we consider the tensorial properties of
the algebra generators in respect to the reduction chain:
\begin{equation}
Sp(12,R)\supset U(6)\supset U(3)\otimes U(2)\supset O(3)\otimes
U(1).  \label{RotLimit}
\end{equation}%
and also classify the basis states  by the quantum numbers
corresponding to the  irreducible representations (irreps) of its
subgroups (Section \ref{s2}). In this way we are able to define
the transition operators between the basis states and then to
evaluate analytically their matrix elements (Section \ref{s3}.).

Transition probabilities are by
definition $SO(3)$ reduced matrix elements of transition operators $T^{E2}$ between the $%
|i\rangle -$initial and $|f\rangle -$final collective states
\begin{equation}
B(E2;L_{i}\rightarrow L_{f})=\frac{1}{2L_{i}+1}\mid \langle \quad f\parallel
T^{E2}\parallel i\quad \rangle \mid ^{2}.  \label{deftrpr}
\end{equation}

As a first step we will test the theory on the transitions between
the states belonging to the ground bands in the even-even nuclei
from the rare earths and the actinides, where the energies and the
staggering between the states are rather well reproduced in our
model approach \cite{Sp12U6}. This proves the correct mapping of
the basis states to the experimentally observed ones and their
band systematic, which is very important for the theoretical
reproduction of the behavior of the physical observables in the
framework of the considered model.

\section{Tensorial properties of the generators of the Sp(12,R) group and
construction of the symplectic basis states of IVBM \label{s2}}

The basic building blocks of the IVBM \cite{IVBM} are the creation
and
annihilation operators of the vector bosons $u_{m}^{\dag}(\alpha )$ and $%
u_{m}(\alpha )$ $(m=0,\pm 1;\alpha =\pm \frac{1}{2}),$ which can
be considered as components of a $6-$dimensional vector, which
transform
according to the fundamental $U(6)$ irreducible representations $%
[1,0,0,0,0,0]_{6}\equiv \lbrack 1]_{6}$ and
$[0,0,0,0,0,-1]_{6}\equiv \lbrack 1]_{6}^{\ast }$, respectively.
These irreducible representations become reducible along the chain
of subgroups (\ref{RotLimit}) defining the dynamical symmetry
\cite{IVBMrl}. This means that along with the quantum number
characterizing the representations of $U(6)$, the operators \ are
also characterized by the quantum numbers of the subgroups of chain (\ref%
{RotLimit}).

The only possible representation of the direct product of
$U(3)\otimes U(2)$ belonging to the representation $[1]_{6}$ of
$U(6)$ is $[1]_{3}.[1]_{2}$, i.e. $[1]_{6}=[1]_{3}.[1]_{2}$.
According to the reduction rules for the decomposition
$U(3)\supset O(3)$ the representation $[1]_{3}$ of $U(3)$ contains
the representation $(1)_{3}$ of the group $O(3)$ giving the
angular momentum of the bosons $l=1$ with a projection $m=0,\pm
1$. The representation $[1]_{2}$ of $U(2)$ defines the
\textquotedblright pseudospin\textquotedblright\ of the bosons
$T=\frac{1}{2}$, whose projection is given by the corresponding
representation of $U(1)$, i. e. $\alpha =\pm \frac{1}{2}$. In this
way the creation and annihilation operators $u_{m}^{\dag}(\alpha
)$ and $u_{m}(\alpha )$are defined as irreducible tensors along
the chain (\ref{RotLimit}) and the used phase convention and
commutation relations are the following \cite{Alisauskas}:
\begin{equation}
\left( u_{[1]_{3}[1]_{2}m\alpha }^{[1]_{6}}\right)
^{\dag}=u_{[1]_{3}^{\ast }[1]_{2}^{\ast }}^{[1]_{6}^{\ast }\quad
m\alpha }=(-1)^{m+\frac{1}{2}-\alpha }u_{[1]_{3}^{\ast
}[1]_{2}^{\ast }-m-\alpha }^{[1]_{6}^{\ast }\quad } \label{bocrao}
\end{equation}

\begin{equation*}
\left[ u_{[1]_{3}^{\ast }[1]_{2}^{\ast }}^{[1]_{6}^{\ast }\quad
m\alpha },u_{[1]_{3}[1]_{2}n\beta }^{[1]_{6}}\right] =\delta
_{m,n}\delta _{\alpha ,\beta }
\end{equation*}

Initially the generators of the symplectic group $Sp(12,R)$ were
written as
double tensors \cite{AGSF} with respect to the $O(3)\supset O(2)$ and $%
U(2)\supset U(1)$ reductions
\begin{equation}
A_{TT_{0}}^{LM}=\sum_{m,n}\sum_{\alpha ,\beta }C_{1m1n}^{LM}C_{\frac{1}{2}%
\alpha \frac{1}{2}\beta }^{TT_{0}}\quad u_{[1]_{3}[1]_{2}m\alpha
}^{[1]_{6}}u_{[1]_{3}^{\ast }[1]_{2}^{\ast }}^{[1]_{6}^{\ast }\quad
\beta n}, \label{A}
\end{equation}%
\begin{equation}
F_{TT_{0}}^{LM}=\sum_{m,n}\sum_{\alpha ,\beta }C_{1m1n}^{LM}C_{\frac{1}{2}%
\alpha \frac{1}{2}\beta }^{TT_{0}}\quad u_{[1]_{3}[1]_{2}m\alpha
}^{[1]_{6}}u_{[1]_{3}[1]_{2}n\beta }^{[1]_{6}},  \label{F}
\end{equation}%
\begin{equation}
G_{TT_{0}}^{LM}=\sum_{m,n}\sum_{\alpha ,\beta }C_{1m1n}^{LM}C_{\frac{1}{2}%
\alpha \frac{1}{2}\beta }^{TT_{0}}\quad u_{[1]_{3}^{\ast
}[1]_{2}^{\ast }}^{[1]_{6}^{\ast }\quad \alpha m}u_{[1]_{3}^{\ast
}[1]_{2}^{\ast }}^{[1]_{6}^{\ast }\quad \beta n}.  \label{G}
\end{equation}%
Further they can be defined as irreducible tensor operators
according to the
whole chain (\ref{RotLimit})\ of subgroups and expressed in terms of (\ref{A}%
), (\ref{F}) and (\ref{G})
\begin{equation}
A_{[\lambda ]_{3}[2T]_{2}\,  TT_{0}}^{\quad \lbrack \chi
]_{6}\quad  LM}=C_{[1]_{3}[1]_{2}[1]_{3}^{\ast }[1]_{2}^{\ast
}\quad \lbrack \lambda ]_{3}[2T]_{2}}^{[1]_{6}\quad \quad \lbrack
1]_{6}^{\ast }\quad \quad \quad \lbrack \chi
]_{6}}C_{(1)_{3}(1)_{3}(L)_{3}}^{[1]_{3}[1]_{3}^{\ast }[\lambda
]_{3}}A_{TT_{0}}^{LM},  \label{Aten}
\end{equation}%
\begin{equation}
F_{[\lambda ]_{3}[2T]_{2}\, TT_{0}}^{\quad \lbrack \chi ]_{6}\quad
 LM}=C_{[1]_{3}[1]_{2}[1]_{3}[1]_{2}\quad \lbrack \lambda
]_{3}[2T]_{2}}^{[1]_{6}\quad \quad \lbrack 1]_{6}\quad \quad \quad
\lbrack \chi
]_{6}}C_{(1)_{3}(1)_{3}(L)_{3}}^{[1]_{3}[1]_{3}[\lambda
]_{3}}F_{TT_{0}}^{LM},  \label{Ften}
\end{equation}%
\begin{equation}
G_{[\lambda ]_{3}[2T]_{2}\, TT_{0}}^{\quad \lbrack \chi ]_{6}\quad
 LM}=C_{[1]_{3}^{\ast }[1]_{2}^{\ast }[1]_{3}^{\ast
}[1]_{2}^{\ast }\quad \lbrack \lambda
]_{3}[2T]_{2}}^{[1]_{6}^{\ast }\quad \quad \lbrack 1]_{6}^{\ast
}\quad \quad \quad \lbrack \chi
]_{6}}C_{(1)_{3}(1)_{3}(L)_{3}}^{[1]_{3}^{\ast }[1]_{3}^{\ast
}[\lambda ]_{3}}G_{TT_{0}}^{LM},  \label{Gten}
\end{equation}%
where, according to the lemma of Racah \cite{Racah}, the
Clebsch-Gordan coefficients along the chain are factorized by
means of the isoscalar factors (IF), defined for each step of
decomposition (\ref{RotLimit}). It should be
pointed out \cite{Alisauskas} that the $U(6)-$ and $U(3)-$IF's, entering in (%
\ref{Aten}), (\ref{Ften}) and (\ref{Gten}), are equal to $\pm $1.

The tensors (\ref{Aten}), transform according to the direct product
$[\chi
]_{6}$ of the corresponding $U(6)$ representations $[1]_{6}$ and $%
[1]_{6}^{\ast }$ \cite{Alisauskas}, namely
\begin{equation}
\lbrack 1]_{6}\times \lbrack 1]_{6}^{\ast }=[1,-1]_{6}+[0]_{6},
\label{prU6A}
\end{equation}%
where $[1,-1]_{6}=[2,1,1,1,1,0]_{6}$ and $[0]_{6}=[1,1,1,1,1,1]_{6}$
is the scalar $U(6)$ representation. Further we multiply the two
conjugated
fundamental representations of $U(3)\otimes U(2)$%
\begin{eqnarray}
&&[1]_{3}[1]_{2}\times \lbrack 1]_{3}^{\ast }[1]_{2}^{\ast }  \notag \\
&=&([1]_{3}\times \lbrack 1]_{3}^{\ast })([1]_{2}\times \lbrack
1]_{2}^{\ast
})  \notag \\
&=&([210]_{3}\oplus \lbrack 1,1,1]_{3})\times ([2,0]_{2}\oplus
\lbrack
1,1]_{2})  \notag \\
&=&[210]_{3}[2]_{2}\oplus \lbrack 210]_{3}[0]_{2}\oplus \lbrack
0]_{3}[2]_{2}\oplus \lbrack 0]_{3}[0]_{2}.  \label{u3u2prd}
\end{eqnarray}%
Obviously the first three $U(3)\otimes U(2)$ irreducible
representations
in the resulting decomposition (\ref{u3u2prd}) belong to the $%
[1,-1]_{6}$ of $U(6)$ and the last one to $[0]_{6}$.

Introducing the notations $u_{i}^{\dag}(\frac{1}{2})=p_{i}^{\dag}$ and $u_{i}^{\dag}(-%
\frac{1}{2})=n_{i}^{\dag}$, the scalar operator
\begin{equation}
A_{[0]_{3}[0]_{2}\ \ 00}^{\ \ \lbrack 0]_{6}\ \ \ 00}=\widehat{N}=%
\frac{1}{\sqrt{2}}\sum_{m}C_{1m1-m}^{00}\left(
p_{m}^{\dag}p_{-m}+n_{m}^{\dag}n_{-m}\right)  \label{Nsc}
\end{equation}%
has the physical meaning of the total number of bosons operator $\widehat{N}%
= $ $\widehat{N_{p}}+$ $\widehat{N_{n}},$ where $\widehat{N_{p}}=\sum $ $%
p_{m}^{\dag}p_{m}$ , $\widehat{N_{n}}=\sum $ $n_{m}^{\dag}n_{m}$
and is obviously the first order invariant of all the unitary
groups $U(6),U(3)$ and $U(2)$. Hence it reduces them to their
respective unimodular subgroups $SU(6),SU(3)$ and $SU(2)$.
Something more, the invariant operator $(-1)^{N}$ , decomposes the
action space $\mathcal{H}$ of the $sp(12,R)$ generators to the
even ${\mathcal{H}}_{+}$ with $N=0,2,4,\ldots,$ and odd
${\mathcal{H}}_{-}$ with $N=1,3,5,...,$ subspaces of the boson
representations of $Sp(12,R)$ \cite{Sp2NRbr}.

In terms of Elliott's notations \cite{Elliott} $\left( \lambda ,\mu
\right) $
where $\lambda =n_{1}-n_{2},\mu =n_{2}-n_{3}$ we have $[210]_{3}=(1,1)$ and $%
[0]_{3}=(0,0)$. The corresponding values of $L$ from the
$SU(3)\supset O(3)$ reduction rules are $L=1,2$ in the $(1,1)$
irrep and $L=0$ in the $(0,0)$.
The values of $T$ are $1$ and $0$ for the $U(2)$ irreps $[2]_{2}$ and $%
[0]_{2}$ respectively. Hence, the $U(2)$ irreps in the direct
product distinguish the equivalent $U(3)$ irreps that appear in this
reduction and there is not degeneracy. The tensors with $T=0$
correspond to the $SU(3)$ generators

\begin{equation}
A_{[210]_{3}[0]_{2}\quad 00}^{\quad \lbrack 1-1]_{6}\quad 1M}=\frac{1}{%
\sqrt{2}}\sum_{m,k}C_{1m1k}^{1M}\left(
p_{m}^{\dag}p_{k}+n_{m}^{\dag}n_{k}\right) \label{Lten}
\end{equation}
\begin{equation}
A_{[210]_{3}[0]_{2}\quad 00}^{\quad \lbrack 1-1]_{6} \quad 2M}=\frac{1}{%
\sqrt{2}}\sum_{m,k}C_{1m1k}^{2M}\left(
p_{m}^{\dag}p_{k}+n_{m}^{\dag}n_{k}\right) \label{Qaten}
\end{equation}
representing the components of the angular $L_{M}$ and Elliott's
quadrupole $Q_{M}$ momenta operators \cite{Elliott}. The tensors
\begin{eqnarray}
A_{[0]_{3}[2]_{2} \quad 11}^{\,[1-1]_{6}\quad 00} &=&\sqrt{\frac{3}{2}}%
\sum_{m}p_{m}^{\dag}n_{-m}\sim T_{1},\notag \\
A_{[0]_{3}[2]_{2}\quad 1-1}^{\, [1-1]_{6}\quad 00} &=&-\sqrt{\frac{3}{2}}%
\sum_{m}n_{m}^{\dag}p_{-m}\sim T_{-1}  \label{Tten}
\end{eqnarray}%
\begin{equation*}
A_{[0]_{3}[2]_{2}\quad 10}^{\, [1-1]_{6}\quad 00}=-\frac{\sqrt{3}}{2}%
\sum_{m}(p_{m}^{\dag}p_{-m}-n_{m}^{\dag}n_{-m})\sim T_{0},
\end{equation*}%
correspond to the $SU(2)$ generators, which are the components of
the pseudospin operator $\widehat{T}$. And finally the tensors
\begin{eqnarray}
A_{[210]_{3}[2]_{2}\quad 11}^{\quad \lbrack 1-1]_{6}\quad LM}
&=&\sum_{m,k}C_{1m1k}^{LM}p_{m}^{+}n_{k},\qquad  \label{A61} \\
A_{[210]_{3}[2]_{2}\quad 1-1}^{\quad \lbrack 1-1]_{6}\quad LM}
&=&\sum_{m,k}C_{1m1k}^{LM}n_{m}^{+}p_{k}  \label{A6-1}
\end{eqnarray}%
and
\begin{equation}
A_{[210]_{3}[2]_{2}\quad 10}^{\quad \lbrack 1-1]_{6}\quad LM}=\frac{1}{\sqrt{%
2}}\sum_{m,k}C_{1m1k}^{LM}(p_{m}^{+}p_{k}-n_{m}^{+}n_{k}),
\label{A60}
\end{equation}%
with $L=1,2$ and $M=-L,-L+1,...,L$ \ extend the $U(3)\otimes U(2)$
algebra to the $U(6)$ one.

By analogy, the tensors (\ref{Ften}) and (\ref{Gten}) transform
according to \cite{Alisauskas}

\begin{equation}
\lbrack 1]_{6}\times \lbrack 1]_{6}=[2]_{6}+[1,1]_{6},
\label{prU6f}
\end{equation}
and

\begin{equation*}
\lbrack 1]_{6}^{\ast }\times \lbrack 1]_{6}^{\ast
}=[-2]_{6}+[-1,-1]_{6},
\end{equation*}%
respectively. But, since the basis states of the IVBM are fully
symmetric, we consider only the fully symmetric $U(6)$
representations $[2]_{6}$ and its conjugated $[-2]_{6}$, since
for  the operators (\ref{Ften}) and (\ref{Gten}) we have
$(F_{[\lambda
]_{3}[2T]_{2}TT_{0}}^{\quad \lbrack \chi ]_{6}\quad LM})^{\dag}$ $%
=(-1)^{\lambda +\mu +L-M+T-T_{0}}$ $G_{[\lambda ]_{3}^{\ast
}[2T]_{2}^{\ast }T-T_{0}}^{\quad \lbrack \chi ]_{6}^{\ast }\quad
L-M}$, where $[\lambda ]_{3}=(\lambda ,\mu )$. Hence we present
the next decompositions only for the $F$ tensors (\ref{prU6f}).
According to the decomposition rules for the fully symmetric
$U(6)$ irreps \cite{Alisauskas} we have

\begin{equation}
\lbrack
2]_{6}=[2]_{3}[2]_{2}+[1,1]_{3}[0]_{2}=(2,0)[2]_{2}+(0,1)[0]_{2}
\label{deU6fsr}
\end{equation}%
which further contain in $(2,0)$ $L=0,2$ with $T=1$ and in $(0,1)$ -
$L=1$
with $T=0.$ Their explicit expressions in terms of the creation $%
p_{i}^{\dag},n_{i}^{\dag}$ and annihilation operators
$p_{i},n_{i}$ at $i=0,\pm 1$ are:
\begin{eqnarray}
F_{[2]_{3}[2]_{2}\quad 1\ 1}^{\quad \lbrack 2]_{6}\quad LM}
&=&\sum_{m,k}C_{1m1k}^{LM}p_{m}^{\dag}p_{k}^{\dag},\qquad  \notag \\
F_{[2]_{3}[2]_{2}\quad 1 -1}^{\quad \lbrack 2]_{6}\quad LM}
&=&\sum_{m,k}C_{1m1k}^{LM}n_{m}^{\dag}n_{k}^{\dag}  \label{F1ex}
\end{eqnarray}%
\begin{equation*}
F_{[2]_{3}[2]_{2}\quad 1\ 0}^{\quad \lbrack 2]_{6}\quad LM}=\frac{1}{\sqrt{2}}%
\sum_{m,k}C_{1m1k}^{LM}\left(p_{m}^{\dag}n_{k}^{\dag}-n_{m}^{\dag}p_{k}^{\dag}\right),
\end{equation*}%
\begin{equation}
F_{[1,1]_{3}[0]_{2}\quad \ 0\ 0}^{\quad \lbrack 2]_{6}\quad \quad LM}=\frac{1}{%
\sqrt{2}}\sum_{m,k}C_{1m1k}^{LM}\left(
p_{m}^{\dag}n_{k}^{\dag}+n_{m}^{\dag}p_{k}^{\dag}\right).
\label{F0ex}
\end{equation}%
In addition to the $SU(3)$ raising generators (\ref{F1ex})
$F_{(2,0)}^{[2]_6}$ we have the operator $F_{(0,1)}^{[2]_6}$
(\ref{F0ex}), which is a new one compared to the generators of the
$Sp(6,R)$ model of Rosensteel and Rowe \cite{RRAnn}.

The above operators and their conjugated ones $G_{[\lambda
]_{3}^{\ast }[2T]_{2}^{\ast }\quad TT_{0}}^{\quad \lbrack \chi
]_{6}^{\ast }\quad \quad LM}$ \ change the number of bosons by two
and realize the symplectic extension of the $U(6)$ algebra. In this
way we have listed all the
irreducible tensor operators in respect to the reduction chain (\ref%
{RotLimit}), that correspond to the infinitesimal operators of the
$Sp(12,R)$ algebra.

Next we can introduce the tensor products

\begin{eqnarray}
&&T_{\qquad \qquad \  \lbrack \lambda ]_{3}[2T]_{2}\quad \ \
TT_{0}}^{([\chi _{1}]_{6}[\chi _{2}]_{6})\quad \omega \lbrack \chi
]_{6}\quad \quad LM}=
\notag \\
&&  \notag \\
&&\sum T_{[\lambda _{1}]_{3}[2T_{1}]_{2}\quad
T_{1}(T_{0})_{1}}^{\quad \lbrack \chi _{1}]_{6}\quad \quad \
L_{1}M_{1}}T_{[\lambda _{2}]_{3}[2T_{2}]_{2}\quad
T_{2}(T_{0})_{2}}^{\quad \lbrack \chi
_{2}]_{6}\quad \quad \ \  L_{2}M_{2}}\times  \notag \\
&&  \notag \\
&&C_{[\lambda _{1}]_{3}[T_{1}]_{2}\ \ [\lambda
_{2}]_{3}[T_{2}]_{2}\quad \lbrack \lambda ]_{3}[2T]_{2}}^{[\chi
_{1}]_{6}\quad \quad \quad \lbrack \chi _{2}]_{6}\quad \quad \
\omega \lbrack \chi ]_{6}}C_{(L_{1})_{3} \ \ (L_{2})_{3}\ \
(L)_{3}}^{[\lambda _{1}]_{3}\ \ [\lambda
_{2}]_{3}\ \ \ [\lambda ]_{3}}\times  \notag \\
&&  \notag \\
&&C_{M_{1}\quad M_{2}\quad M}^{L_{1}\quad L_{2}\quad \
L}C_{(T_{0})_{1} \ \ (T_{0})_{2}\quad T_{0}}^{T_{1}\qquad
T_{2}\qquad T} \label{tpr}
\end{eqnarray}%
of two tensor operators $T_{[\lambda ]_{3}[2T]_{2}\quad
TT_{0}}^{\quad \lbrack \chi ]_{6}\quad \quad LM},$ which are as
well tensors in respect to the considered reduction chain. We use
(\ref{tpr}) to obtain the tensorial properties of the operators in
the enveloping algebra of $Sp(12,R),$ containing the products of
the algebra generators. In this particular case we are interested
in the transition operators between states differing by four
bosons $T_{[\lambda ]_{3}[2T]_{2}\quad TT_{0}}^{\quad[4]_{6}
\qquad LM}$, expressed in terms of the products of two operators $%
F_{[\lambda ]_{3}[2T]_{2}\quad TT_{0}}^{\quad \lbrack 2]_{6}\quad \quad LM}$%
. Making use of the decomposition (\ref{deU6fsr}) \ and the
reduction rules in the chain (\ref{RotLimit}), we list in Table 1
all the representations of the chain subgroups that define the
transformation properties of the resulting tensors.

\begin{table}[h]
\caption{Tensor products of two raising operators.} \label{T1}
\smallskip \centering
\begin{tabular}{llllll}
\\ \hline
\multicolumn{1}{|l}{%
\begin{tabular}{l}
$\lbrack 2]_{6}$ \\
$\lbrack \lambda _{1}]_{3}[2T_{1}]_{2}$%
\end{tabular}%
} & \multicolumn{1}{|l}{%
\begin{tabular}{l}
$\lbrack 2]_{6}$ \\
$\lbrack \lambda _{2}]_{3}[2T_{1}]_{2}$%
\end{tabular}%
} & \multicolumn{1}{|l}{%
\begin{tabular}{l}
$\lbrack 4]_{6}$ \\
$\lbrack \lambda ]_{3}[2T]_{2}$%
\end{tabular}%
} & \multicolumn{1}{|l}{%
\begin{tabular}{l}
$O(3)$ \\
K; $L$%
\end{tabular}%
} & \multicolumn{1}{|l}{%
\begin{tabular}{l}
$U(2)$ \\
$T$%
\end{tabular}%
} & \multicolumn{1}{|l|}{%
\begin{tabular}{l}
$U(1)$ \\
$T_{0}$%
\end{tabular}%
} \\ \hline \multicolumn{1}{|l}{$(2,0)[2]_{2}$} &
\multicolumn{1}{|l}{$(2,0)[2]_{2}$} &
\multicolumn{1}{|l}{$(4,0)[4]_{2}$} & \multicolumn{1}{|l}{0;$0,2,4$}
& \multicolumn{1}{|l}{$2$} & \multicolumn{1}{|l|}{$0,\pm 1,\pm 2$}
\\ \hline \multicolumn{1}{|l}{$(2,0)[2]_{2}$} &
\multicolumn{1}{|l}{$(2,0)[2]_{2}$} &
\multicolumn{1}{|l}{$(2,1)[2]_{2}$} & \multicolumn{1}{|l}{1;$1,2,3$}
& \multicolumn{1}{|l}{$1$} & \multicolumn{1}{|l|}{$0,\pm 1$} \\
\hline \multicolumn{1}{|l}{$(2,0)[2]_{2}$} &
\multicolumn{1}{|l}{$(2,0)[2]_{2}$} &
\multicolumn{1}{|l}{$(0,2)[0]_{2}$} & \multicolumn{1}{|l}{0;$0,2$} &
\multicolumn{1}{|l}{$0$} & \multicolumn{1}{|l|}{$0$} \\ \hline
\multicolumn{1}{|l}{$(2,0)[2]_{2}$} &
\multicolumn{1}{|l}{$(0,1)[0]_{2}$} &
\multicolumn{1}{|l}{$(2,1)[2]_{2}$} & \multicolumn{1}{|l}{1;$1,2,3$}
& \multicolumn{1}{|l}{$1$} & \multicolumn{1}{|l|}{$0,\pm 1$} \\
\hline \multicolumn{1}{|l}{$(0,1)[0]_{2}$} &
\multicolumn{1}{|l}{$(0,1)[0]_{2}$} &
\multicolumn{1}{|l}{$(0,2)[0]_{2}$} & \multicolumn{1}{|l}{0;$0,2$} &
\multicolumn{1}{|l}{$0$} & \multicolumn{1}{|l|}{$0$} \\ \hline
\end{tabular}%
\end{table}

In order to clarify the role of the tensor operators introduced in
this section as transition operators and to simplify the
calculation of their matrix elements, the basis for the Hilbert
space must be symmetry adapted to the  algebraic structure along
the considered subgroup chain (\ref{RotLimit}). It is evident from
(\ref{F1ex}) and (\ref{F0ex}), that the basis states of the IVBM
in the $\mathcal{H}_{+}$ \ ($N-$even)  subspace of the boson
representations of $\ Sp(12,R)$ \ can be obtained by a consecutive
application of the raising operators $F_{[\lambda
]_{3}[2T]_{2}\quad TT_{0}}^{\quad \lbrack 2]_{6}\quad \quad LM}$ \
on the boson vacuum $\mid 0 \ \rangle$ (ground state)
, annihilated by the tensor operators $%
G_{[\lambda ]_{3}[2T]_{2}\quad TT_{0}}^{\quad \lbrack \chi
]_{6}\quad \quad LM}$ $\mid 0 \ \rangle =0$ \ and $A_{[\lambda
]_{3}[2T]_{2}\quad TT_{0}}^{\quad \lbrack \chi ]_{6}\quad \quad
LM}\mid 0 \ \rangle =0$.

Thus, in general a basis for the considered dynamical symmetry of
the IVBM can be constructed by applying the multiple symmetric
coupling (\ref{tpr}) of the raising tensors $F_{[\lambda
_{i}]_{3}[2T_{i}]_{2}\quad
T_{i}T_{0i}}^{\quad \lbrack 2]_{6}\quad \quad L_{i}M_{i}}$ with itself \ - $%
[F\times \ldots \times \quad F]_{[\lambda ]_{3}[2T]_{2}\quad
TT_{0}}^{\quad \lbrack \chi ]_{6}\quad \quad LM}$ . Note that only
fully symmetric tensor products $[\chi ]_{6}\equiv \lbrack N]_{6}$ \
are nonzero, since the raising operator commutes with itself. The
possible $U(3)$ couplings are enumerated by the set $[\lambda
]_{3}=\{[n_{1},n_{2},0]\equiv (\lambda = n_{1}-n_{2},\mu =
n_{2});n_{1}\geq n_{2}\geq 0$ $\}$. The number of
copies of the operator $F$ in the symmetric product tensor $[N]_{6}$ is $%
N/2$, where $N=\lambda +2\mu $ \cite{Sp12U6}. Each raising
operator will increase the number of bosons $N$ by two. Then, the
resulting infinite basis is denoted by:
 \begin{equation}
 |[N](\lambda ,\mu);KLM;TT_{0}\  \rangle  , \label{bast}
 \end{equation}
 where $KLM$ are the quantum numbers for
the non-orthonormal basis of the irrep $(\lambda ,\mu )$.

The $Sp(12,R)$ classification scheme for the $SU(3)$ boson
representations obtained by applying the reduction rules
\cite{Sp12U6} for the irreps in the chain (\ref{RotLimit}) for even
value of the number of bosons $N$ \ is shown on Table \ref{T2}. Each
row (fixed $N$) of the table corresponds to a given irreducible
representation of the $U(6)$ algebra. Then the possible values for
the pseudospin, given in the column next to the respective value of
$N$,
are $T=\frac{N}{2},\frac{N}{2}-1,...$ $0$. Thus when $N$ and $T$ are fixed, $%
2T+1$ equivalent representations of the group $SU(3)$ arise. Each of
them is distinguished by the eigenvalues of the operator
$T_{0}:-T,-T+1,...,T,$
defining the columns of Table \ref{T2}. The same $SU(3)$ representations $%
(\lambda ,\mu )$ arise for the positive and negative eigenvalues of
$T_{0}$.

\begin{table}[h]
\caption{Classification of the basis states.} \label{T2}
\smallskip \centering%
\begin{tabular}{||l||rrrr|l|}
\hline\hline $N$ T & $T_{_{0}}\backslash $ $...\ \pm 4$ &
\multicolumn{1}{||r}{$\pm 3$} & \multicolumn{1}{||r}{$\pm 2$} &
\multicolumn{1}{||r|}{$\ \pm 1$} & \multicolumn{1}{||l||}{$\ \ 0$}
\\ \hline\hline
\multicolumn{1}{||r||}{$0%
\begin{tabular}{l}
$0$%
\end{tabular}%
$} & \multicolumn{1}{||c}{} & \multicolumn{1}{l}{} &
\multicolumn{1}{l}{} & \multicolumn{1}{c|}{$\swarrow
F_{[2]_{3}[2]_{2}\quad }^{\quad \lbrack 2]_{6}\quad }$} &
\begin{tabular}{l}
$(0,0)$%
\end{tabular}
\\ \cline{1-1}\cline{5-6}
\multicolumn{1}{||r||}{$2%
\begin{tabular}{l}
$1$ \\
$0$%
\end{tabular}%
$} & \multicolumn{1}{||c}{} & \multicolumn{1}{l}{} & \multicolumn{1}{l}{$%
F_{[1,1]_{3}[0]_{2}}^{\quad \lbrack 2]_{6}\quad }\downarrow $} &
\multicolumn{1}{|r|}{$%
\begin{tabular}{c}
$\Longrightarrow $ \\
\multicolumn{1}{l}{$A_{[2,1]_{3}[0]_{2}}^{\, \lbrack 1-1]_{6}}$}%
\end{tabular}%
$%
\begin{tabular}{l}
$(2,0)$ \\
\multicolumn{1}{c}{$-$}%
\end{tabular}%
} &
\begin{tabular}{l}
$(2,0)$ \\
$(0,1)$%
\end{tabular}
\\ \cline{1-1}\cline{4-6}
\multicolumn{1}{||r||}{$4%
\begin{tabular}{l}
$2$ \\
$1$ \\
$0$%
\end{tabular}%
$} & \multicolumn{1}{||c}{} & \multicolumn{1}{l}{} & \multicolumn{1}{|r}{%
\begin{tabular}{l}
$(4,0)$ \\
\multicolumn{1}{c}{$-$} \\
\multicolumn{1}{c}{$-$}%
\end{tabular}%
} & \multicolumn{1}{|r|}{$A_{[2,1]_{3}[2]_{2}}^{\, \lbrack
1-1]_{6}}\Downarrow $\
\begin{tabular}{l}
$(4,0)$ \\
$(2,1)$ \\
\multicolumn{1}{c}{$-$}%
\end{tabular}%
} &
\begin{tabular}{l}
$(4,0)$ \\
$(2,1)$ \\
$(0,2)$%
\end{tabular}
\\ \cline{1-1}\cline{3-5}\cline{5-6}
\multicolumn{1}{||r||}{$6%
\begin{tabular}{l}
$3$ \\
$2$ \\
$1$ \\
$0$%
\end{tabular}%
$} & \multicolumn{1}{||c}{%
\begin{tabular}{l}
$A_{[0]_{3}[2]_{2}}^{\, \lbrack 1-1]_{6}}$ \\
\multicolumn{1}{c}{$\rightarrow $}%
\end{tabular}%
} & \multicolumn{1}{|r}{%
\begin{tabular}{l}
$(6,0)$ \\
\multicolumn{1}{c}{$-$} \\
\multicolumn{1}{c}{$-$} \\
\multicolumn{1}{c}{$-$}%
\end{tabular}%
} & \multicolumn{1}{|r}{%
\begin{tabular}{l}
$(6,0)$ \\
$(4,1)$ \\
\multicolumn{1}{c}{$-$} \\
\multicolumn{1}{c}{$-$}%
\end{tabular}%
} & \multicolumn{1}{|r|}{%
\begin{tabular}{l}
$(6,0)$ \\
$(4,1)$ \\
$(2,2)$ \\
\multicolumn{1}{c}{$-$}%
\end{tabular}%
} &
\begin{tabular}{l}
$(6,0)$ \\
$(4,1)$ \\
$(2,2)$ \\
$(0,3)$%
\end{tabular}
\\ \cline{1-2}\cline{2-6}
\multicolumn{1}{||r||}{$8%
\begin{tabular}{l}
$4$ \\
$3$ \\
$2$ \\
$1$ \\
$0$%
\end{tabular}%
$} &
\begin{tabular}{l}
$(8,0)$ \\
\multicolumn{1}{c}{$-$} \\
\multicolumn{1}{c}{$-$} \\
\multicolumn{1}{c}{$-$} \\
\multicolumn{1}{c}{$-$}%
\end{tabular}
& \multicolumn{1}{|r}{%
\begin{tabular}{l}
$(8,0)$ \\
$(6,1)$ \\
\multicolumn{1}{c}{$-$} \\
\multicolumn{1}{c}{$-$} \\
\multicolumn{1}{c}{$-$}%
\end{tabular}%
} & \multicolumn{1}{|r}{$\ \ $%
\begin{tabular}{l}
$(8,0)$ \\
$(6,1)$ \\
$(4,2)$ \\
\multicolumn{1}{c}{$-$} \\
\multicolumn{1}{c}{$-$}%
\end{tabular}%
} & \multicolumn{1}{|r|}{%
\begin{tabular}{l}
$(8,0)$ \\
$(6,1)$ \\
$(4,2)$ \\
$(2,3)$ \\
\multicolumn{1}{c}{$-$}%
\end{tabular}%
} &
\begin{tabular}{l}
$(8,0)$ \\
$(6,1)$ \\
$(4,2)$ \\
$(2,3)$ \\
$(0,4)$%
\end{tabular}
\\ \cline{1-1}\cline{2-2}\cline{2-6}
\multicolumn{1}{||r||}{$...$} & $...$ & $...$ &
\multicolumn{1}{|r}{$...$} &
\multicolumn{1}{|r|}{$...$} & \multicolumn{1}{|r|}{$...$}%
\end{tabular}%
\end{table}

Now it is clear which of the tensor operators act as transition
operators between the basis states ordered in the classification
scheme presented on Table \ref{T2}. The operators $F_{[\lambda
]_{3}[2T]_{2}\quad TT_{0}}^{\quad \lbrack 2]_{6}\quad \quad LM}$
with $T_{0}=0$ (\ref{F0ex}) give the transitions between two
neighboring cells $(\downarrow )$ from one column,
while the ones with $T_{0}=\pm 1$(\ref{F1ex}) change the column as well $%
(\swarrow )$. The \ tensors $A_{[2,1]_{3}[0]_{2}}^{\quad \lbrack 1-1]_{6}}$ (%
\ref{Lten}) and (\ref{Qaten}), which correspond to the $SU(3)$
generators do not change the $SU(3)$ representations $(\lambda
,\mu ),$ but can change the angular momentum $L$ inside it
$(\Longrightarrow )$. The $SU(2)$ generating tensors
$A_{[0]_{3}[2]_{2}}^{ \lbrack 1-1]_{6}}$(\ref{Tten}) change the
projection $T_{_{0}}(\rightarrow )$ of the pseudospin $T$ and in
this way distinguish the equivalent $SU(3)$ irreps belonging to
the different columns of the same row of Table \ref{T2}. Inside a
given cell the transition between the different $SU(3)$ irreps
$(\Downarrow )$ is realized by the
operators $A_{[2,1]_{3}[2]_{2}}^{\quad \lbrack 1-1]_{6}}$ (\ref{A61}), (\ref%
{A6-1}) and (\ref{A60}), that represent the $U(6)$ generators. The
action of the tensor operators on the $SU(3)$ vectors inside a
given cell or between the cells of Table \ref{T2}. is also
schematically presented on it with corresponding arrows, given
above in parentheses.

\section{Matrix elements of the transition operators in symmetry-adapted
basis \label{s3}}

Physical applications are based on the correspondence of sequences
of $SU(3)$ vectors to sequences of collective states belonging to
different bands in the nuclear spectra. The above analysis,
permits the  definition of the appropriate transition operators as
crresponding combinations of the tensor operators given in Section
\ref{s2}.

Matrix elements of the $Sp(12,R)$ algebra can be calculated in
several ways. A direct method is to use the $Sp(12,R)$ commutation
relations \cite{IVBM} to derive recursion relations. Another is to
start from approximate matrix element and proceed by successive
approximations, adjusting the matrix elements until the commutation
relations are precisely satisfied \cite{RRmatr}. The third method is
to make use of a vector-valued coherent-state representation theory
\cite{AGSF},\cite{VCSforSp} to relate the matrix elements to the
known matrix elements of a much simpler Weyl algebra.

However, in the present paper we use another technique for
calculation of the matrix elements of the $Sp(12,R)$ algebra,
based on the fact that the representations of the $SU(3)$ subgroup
in IVBM are build with the help of the two kind of vector bosons,
which is in some sense simpler than the construction of the
$SU(3)$ representations in IBM and $Sp(6,R)$ symplectic model.

In the preceding section we expressed the $Sp(12,R)$ generators $%
F_{TT_{0}}^{LM},$ $G_{TT_{0}}^{LM},$ $A_{TT_{0}}^{LM}$ and the
basis states as components of irreducible tensors in respect to
the reduction chain (\ref{RotLimit}). Thus, for calculating their
matrix elements, we have the
advantage of using the Wigner-Eckart theorem in two steps. For the $%
SU(3)\rightarrow SO(3)$ and $SU(2)\rightarrow U(1)$ reduction we
need the standard $SU(2)$ Clebsch-Gordan coefficient (CGC)
\begin{widetext}
\begin{equation}
\begin{tabular}{l}
$\langle \lbrack N^{\prime }]\, (\lambda ^{\prime },\mu ^{\prime
});K^{\prime }L^{\prime }M^{\prime };T^{\prime }T_{0}^{\prime
}|T_{[\sigma ]_{3}[2t]_{2}\quad tt_{0}}^{\quad \lbrack \chi
]_{6}\, \quad lm}|[N]\,
(\lambda ,\mu );KLM;TT_{0}\,\rangle $ \\
\\
$=\langle \lbrack N^{\prime }](\lambda ^{\prime },\mu ^{\prime
});K^{\prime }L^{\prime }||T_{[\sigma ]_{3}[2t]_{2}\quad
tt_{0}}^{\quad \lbrack \chi ]_{6}\,\quad lm}||[N](\lambda ,\mu
);KL\rangle C_{LMlm}^{L^{\prime
}M^{\prime }}C_{TT_{0}tt_{0}}^{T^{\prime }T_{0}^{\prime }}.$%
\end{tabular}
 \label{ME}
\end{equation}

For the calculation of the double-barred reduced matrix elements in (\ref{ME}%
) we use the next step:
\begin{equation}
\begin{tabular}{l}
$\langle \lbrack N^{\prime }]\, (\lambda ^{\prime },\mu ^{\prime
});K^{\prime }L^{\prime }||T_{[\sigma ]_{3}[2t]_{2}\quad
tt_{0}}^{\quad \lbrack \chi ]_{6}\quad \quad lm}||[N]\,(\lambda
,\mu );KL\rangle $ \\
\\
$=\langle \lbrack N^{\prime }]|||T_{[\sigma ]_{3}[2t]_{2}}^{\quad
\lbrack \chi ]_{6}}|||[N]\rangle C_{(\lambda ,\mu )[2T]_{2}\quad
\lbrack \sigma ]_{3}[2t]_{2}\quad (\lambda ^{\prime },\mu ^{\prime
})[2T^{\prime }]_{2}}^{[N]_{6\ }\ \ \ \ \ \ \ \ \ \ \ [\chi ]_{6}\ \
\ \ \ \ \ \ \ \ [N^{\prime }]_{6\ }}C_{KL\quad k(l)_{3}\quad
K^{\prime }L^{\prime
}}^{(\lambda ,\mu ) \quad [\lambda ]_{3} \quad (\lambda ^{\prime },\mu ^{\prime })}$,%
\end{tabular}
\label{3-barredME}
\end{equation}
\end{widetext}
where $C_{(\lambda ,\mu )[2T]_{2}\quad \lbrack \sigma
]_{3}[2t]_{2}\quad (\lambda ^{\prime },\mu ^{\prime })[2T^{\prime
}]_{2}}^{[N]_{6\ }\ \ \ \ \ \ \ \ \ \ \ [\chi ]_{6}\ \ \ \ \ \ \ \
\ \ [N^{\prime }]_{6\ }}$ and $C_{KL\quad k(l)_{3}\quad K^{\prime
}L^{\prime }}^{(\lambda ,\mu )\quad[\lambda ]_{3}\quad (\lambda
^{\prime },\mu ^{\prime })}$ are $U(6)$ and $SU(3)$ IF's.
Obviously the practical value of the application of the
generalized Wigner-Eckart theorem for the calculation of the
matrix elements of the $Sp(12,R)$ generators and the construction
of the symplectic basis depend on the knowledge of the isoscalar
factors for the reductions $U(6)\supset U(3)\otimes U(2)$ and
$U(3) \supset O(3)$, respectively. For the evaluation of the
matrix elements (\ref{ME}) of the $Sp(12,R)$ operators in respect
to the chain (\ref{RotLimit}) the  reduced triple-barred $U(6)$
matrix elements are also required (\ref{3-barredME}).

\section{B(E2) transition probabilities for the ground  band \label{s4}}

In the symplectic extension of the IVBM the complete spectrum of
the system is obtained in all the even subspaces with fixed $N$-
even of the UIR $[N]_{6}$ of $U(6)$, belonging to a given even UIR
of $Sp(12,R)$. The classification scheme of the $SU(3)$ boson
representations for even values of the number of bosons $N$ was
presented on Table \ref{T2}. The equivalent use of the $(\lambda
,\mu )$ labels, resulting from the connections  $T=\lambda/2$ and
$N=\lambda + 2\mu$ facilitates the final reduction to the $SO(3)$
representations, which define the angular momentum $L$ and its
projection $M.$ The multiplicity index $K$ appearing in this
reduction is related to the projection of $L$ on the body fixed
frame and is used with the parity ($\pi $)\ to label the different
bands ($K^{\pi } $) in the energy spectra of the nuclei. We have
defined the parity of the states as $\pi =(-1)^{T}$ \cite{Sp12U6}.
This allowed us to describe both positive and negative bands.

In this paper we give as an example the evaluation of the  E2
transition probabilities of the ground state band (GSB) \cite{Sp12U6},
whose states were identified  with the $SU(3)$ multiplets $%
(0,\mu )$.  In terms of $(N,T)$ this choice corresponds to
$(N=2\mu ,T=0)$.
 We define the
energies of each state with given $L$ as yrast energy with respect to $%
N$ in the  considered bands. Hence for the ground band their
minimum values are obtained at $N=2L$. Using the tensorial
properties of the $Sp(12,R)$ generators it is easy to define the
$E2$ transition operator between the states of the  considered
band:
\begin{widetext}
\begin{equation}
T^{E2}=e\left[A_{[210]_{3}[0]_{2}\quad 00}^{ \lbrack 1-1]_{6}\quad
\quad 20}+\theta ([F\times F]_{(0,2)[0]_{2}\quad 00}^{\quad
\lbrack 4]_{6}\quad \, \ 20}+[G\times G]_{(2,0)[0]_{2}\quad
00}^{\quad \lbrack -4]_{6}\quad \,20}\right],  \label{te2}
\end{equation}%
\end{widetext}
where the first tensor operator is expressed in terms of the boson creation $%
p_{m}^{\dag},n_{m}^{\dag}$ and annihilation $p_{m},n_{m}$, $m=\pm
1$ operators in (\ref{Qaten}) and as part of the $SU(3)$
generators  actually changes only the angular momentum with
$\Delta L=2$.

The tensor product
\begin{widetext}
\begin{eqnarray} \label{FF}
\lbrack F\times F]_{(0,2)[0]_{2}\quad \, 00}^{\quad \lbrack
4]_{6}\quad \quad 20} &=&\sum C_{(2,0)[2]_{2} \, (2,0)[2]_{2}\quad
(0,2)[0]_{2}}^{\quad [2]_{6}\quad \quad \lbrack 2]_{6}\quad \quad
\lbrack
4]_{6}}C_{\,(2)_{3}\quad (2)_{3}\quad (2)_{3}}^{(2,0)\,\,(2,0)\quad(0,2)} \notag \\
&& \label{FF} \\
&&\times C_{20\ 20}^{20}C_{1 1\ 1 -1}^{10}\ F_{(2,0)[2]_{2}\ \
11}^{\quad \lbrack 2]_{6}\quad \ 20}F_{(2,0)[2]_{2}\ \ 1-1}^{\quad
\lbrack 2]_{6}\quad \ 20} \notag
\end{eqnarray}%
\end{widetext}
of the operators (\ref{F1ex}) that are the pair
raising $Sp(12,R)$ generators changes the number of bosons by
$\Delta N=4$ and $\Delta L=2$. Thus, for calculating the matrix
elements of (\ref{te2}) between the basis states (\ref{bast}), we
have the advantage of using the Wigner-Eckart theorem in two steps
(\ref{ME}) and (\ref{3-barredME}), where  only their reduced
triple-barred $U(6)$ matrix elements are required.

However the $SU(3)$ generators (\ref{Lten}) and (\ref{Qaten}) are
scalars with respect to the isospin group $U(2)$, so they act only
on the $SU(3)$ part of the wave function and the Wigner-Eckart
theorem is applied in respect to the $SU(3)$ subgroup \cite{U6IF}
\begin{widetext}
\begin{equation*}
\begin{tabular}{l}
$\langle \lbrack N],(\lambda ^{\prime },\mu ^{\prime });K^{\prime
}L^{\prime }M^{\prime };T^{\prime }T_{0}^{\prime }|\
A_{(1,1)[0]_{2}\ \ \ 00}^{\ \ [1,-1]_{6}\
\ \ \ lm}\ |[N],(\lambda ,\mu );KLM;TT_{0}\rangle $ \\
\\
$=\delta _{TT^{\prime }}\delta _{T_{0}T_{0}^{\prime }}\delta
_{\lambda \lambda ^{\prime }}\delta _{\mu \mu ^{\prime
}}\sum\limits_{\rho =1,2}C_{K(L)\ \ \ k(l)\ \ K^{\prime
}(L^{\prime })}^{(\lambda ,\mu )\ \ (1,1)\ \
\rho (\lambda ^{\prime },\mu ^{\prime })}$ \\
\\
$\times C_{LM\ \ \ lm}^{L^{\prime }M^{\prime }}\langle \lbrack
N],(\lambda ^{\prime },\mu ^{\prime })|||\
A_{(1,1)[0]_{2}}^{[1,-1]_{6}\ }\
|||[N],(\lambda ,\mu )\rangle.$ \\
\end{tabular}%
\end{equation*}%
\end{widetext}
The sum over $\rho$ runs over terms containing
products of IF's of $SU(3)$ and $U(6)$, respectively. The reduced
triple-barred matrix elements are well known and are given for
$\rho=1$ by \cite{RRAnn}
\begin{equation}
\langle \lbrack N],(\lambda ,\mu )|||A_{(1,1)_{3}[0]_{2}\quad }^{
\lbrack 1-1]_{6}}|||[N],(\lambda ,\mu )\rangle _{1} = \left\{
\begin{array}{c}
g_{\lambda \mu }, \ \ \mu =0 \\
-g_{\lambda \mu }, \  \mu \neq 0%
\end{array}
\right.
\label{RTME of A}
\end{equation}
where
\begin{equation}
g_{\lambda \mu }=2\left(\frac{\lambda ^{2}+\mu ^{2}+\lambda \mu +3\lambda +3\mu }{%
3}\right)^{1/2}
\end{equation}
and the phase convention is chosen to agree with that of Draayer
and Akiyama \cite{DrAk}. For $\rho=2$ we have $\langle \lbrack
N],(\lambda ,\mu )|||A_{[210]_{3}[0]_{2}\quad }^{\quad \lbrack
1-1]_{6}}|||[N],(\lambda ,\mu )\rangle _{2}=0 $.
 Thus, for the matrix elements of
$A_{[210]_{3}[0]_{2}\quad 00}^{\quad \lbrack 1-1]_{6}\quad 20}$
between the states attributed to the GSB we obtain
\begin{widetext}
\begin{equation}
\begin{tabular}{l}
$\langle \lbrack N],(0,\mu );0L-20;00|A_{(1,1)\ [0]_{2}\quad
00}^{\quad
\lbrack 1-1]_{6}\quad  20}|[N],(0,\mu );0L0;00\rangle $ \\
\\
$=C_{L-2\ \quad \ 2\ \quad L}^{(0 ,\mu )\ \ (1,1)\ \ \ (0,\mu)}
C_{L-2,0\ \ 2,0}^{L,0}\langle \lbrack N],(0,\mu )|||A_{(1,1)\
[0]_{2}}^{\quad \lbrack
1-1]_{6}}|||[N],(0,\mu )\rangle $ \\
\\
$=2\left[\dfrac{(\mu -L+2)(\mu
+L+1)(L-1)L}{2(2L-1)(2L+1)}\right]^{1/2}
C_{L-2,0\ \ 2, 0}^{L,0}$%
\end{tabular}
\end{equation}%
The value of the reduced $SU(3)$ Clebsch-Gordan coefficient (IF)
is taken from Ref.\cite{Ver}. Actually, we are interested in the
$SO(3)$ reduced matrix elements which enter in (\ref{deftrpr}).
Thus taking into account the yrast conditions $\mu=L$ we obtain
\begin{equation}
\langle \lbrack N],(0,\mu );0L-2;00||A_{(1,1)\ [0]_{2}}^{\quad
\lbrack 1-1]_{6}}||[N],(0,\mu );0L;00\rangle  = 2 \left[
\frac{(L-1)L}{(2L-1)}\right]^{1/2}. \label{RME of A}
\end{equation}
For the calculation of the matrix element
\begin{equation}
\begin{tabular}{l}
$\langle \lbrack N+4],(0,\mu +2);0L+20;00|[F\times
F]_{(0,2)[0]_{2}\quad
00}^{\quad \lbrack 4]_{6}\quad \quad 20}|[N],(0,\mu );0L0;00\rangle $ \\
\\
$=C_{(0,\mu )[0]_{2}\quad (0,2)[0]_{2}\quad (0,\mu
+2)[0]_{2}}^{[N]_{6\ }\ \ \ \ \ \ \ [4]_{6}\ \ \ \ \ \ \ \
[N+4]_{6\ }} C_{\ \ L\ \quad \ 2\ \quad \ \ \ L+2}^{(0 ,\mu )\ \
(2,0)\ \ \ (0,\mu+2)}
C_{L,0\ \ \ \ \ 2,0}^{L+2,0}$ \\
\\
$\times \langle \lbrack N+4],(0,\mu +2)|||[F\times
F]_{(0,2)[0]_{2}}^{\quad
\lbrack 4]_{6}\quad }|||[N],(0,\mu )\rangle $%
\end{tabular}%
\label{RMEFF}
\end{equation}%
\end{widetext}
we use the standard recoupling technique for two
coupled $U(6)$ tensors \cite{Recoupling}:
\begin{equation}
\begin{tabular}{l}
$\langle \lbrack N^{\prime }]|||\ [T^{[\alpha ]_{6}}\times T^{[\beta
]_{6}}]^{\sigma \lbrack \gamma ]_{6}}\ |||[N]\rangle $ \\
\\
$=\underset{c,\rho _{1},\rho _{2}}{\sum }U([N]_{6};[\beta
]_{6};[N^{\prime }]_{6};[\alpha ]_{6}|[N_{c}]_{6}\rho _{2}\rho
_{1};[\gamma ]_{6}\ \sigma )$
\\
\\
$\times \langle \lbrack N^{\prime }]|||\ T^{[\alpha ]_{6}}\
|||[N_{c}]\rangle \langle \lbrack N_{c}]|||T^{[\beta ]_{6}}\ |||[N]\rangle ,$%
\end{tabular}
\label{RCF}
\end{equation}%
where $U(...)$ are the $U(6)$ Racah coefficients in unitary form
\cite{UNRacah}. For the reduced triple-bared matrix element in our
case, which is multiplicity free and hence there is no sum, we
have
\begin{equation*}
\begin{tabular}{l}
$\langle \lbrack N+4]|||[F\times F]_{(0,2)[0]_{2}}^{\quad \lbrack
4]_{6}\quad }|||[N]\rangle $ \\
\\
$=U([N]_{6};[2]_{6};[N+4]_{6};[2]_{6}|[N+2]_{6};[4]_{6})$ \\
\\
$\times \langle \lbrack N+4]|||\ F^{(2,0)}\ |||[N+2]\rangle \langle
\lbrack
N+2]|||\ F^{(2,0)}\ |||[N]\rangle ,$%
\end{tabular}%
\end{equation*}
where the corresponding Racah coefficient for maximal coupling
representations is equal to unity (\cite{Recoupling}; see also
formula $A9$ of Ref. \cite{UNRacah}). Applying again the formula
(\ref{RCF}) with respect to coupled tensor $F^{[2]_{6}}$ and using
the fact that in the case of vector bosons which span the
fundamental irrep $[1]$ of $u(n)$ algebra the $u(n)$-reduced
matrix element of raising generators has the well known form
\cite{LeRo}
\begin{equation}
\langle \lbrack N+1]|||\ u_{m}^{\dagger }(\alpha )\ |||[N]\rangle
=\sqrt{N+1}. \label{RTME of u}
\end{equation}
we obtain
\begin{equation*}
\begin{tabular}{l}
$\langle \lbrack N+2]|||F^{[2]_{6}}|||[N]\rangle $ \\
\\
$=U([N]_{6};[1]_{6};[N+2]_{6};[1]_{6}|[N+1]_{6};[2]_{6})$ \\
\\
$\times \langle \lbrack N+2]|||\ p^{\dagger \ [1]_{6}}\
|||[N+1]\rangle
\langle \lbrack N+1]|||\ p^{\dagger \ [1]_{6}}\ |||[N]\rangle $ \\
\\
$=\sqrt{(N+1)(N+2)}$%
\end{tabular}%
\end{equation*}%
and in analogy
\begin{equation*}
\langle \lbrack N+4]|||F^{[2]_{6}}|||[N+2]\rangle
=\sqrt{(N+3)(N+4)}.
\end{equation*}%

\begin{figure}[b!]
\centerline{\epsfxsize=3.5in\epsfbox{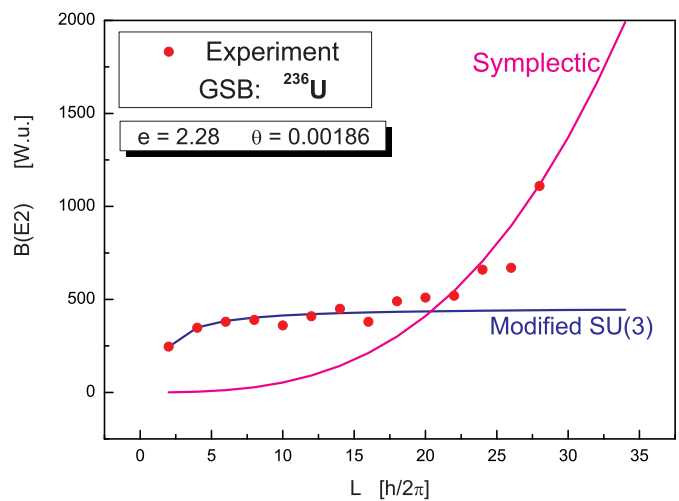}}
% \centering
%\includegraphics[height=8cm]{Contributions.eps}
\caption{(Color online) The behavior of the number conserving and
symplectic terms of the matrix elements of the transition operator
$T^{E2}$(\protect\ref{te2}).}  \label{trbe}
\end{figure}

\begin{figure}[t!]
\centerline{\epsfxsize=3.5in\epsfbox{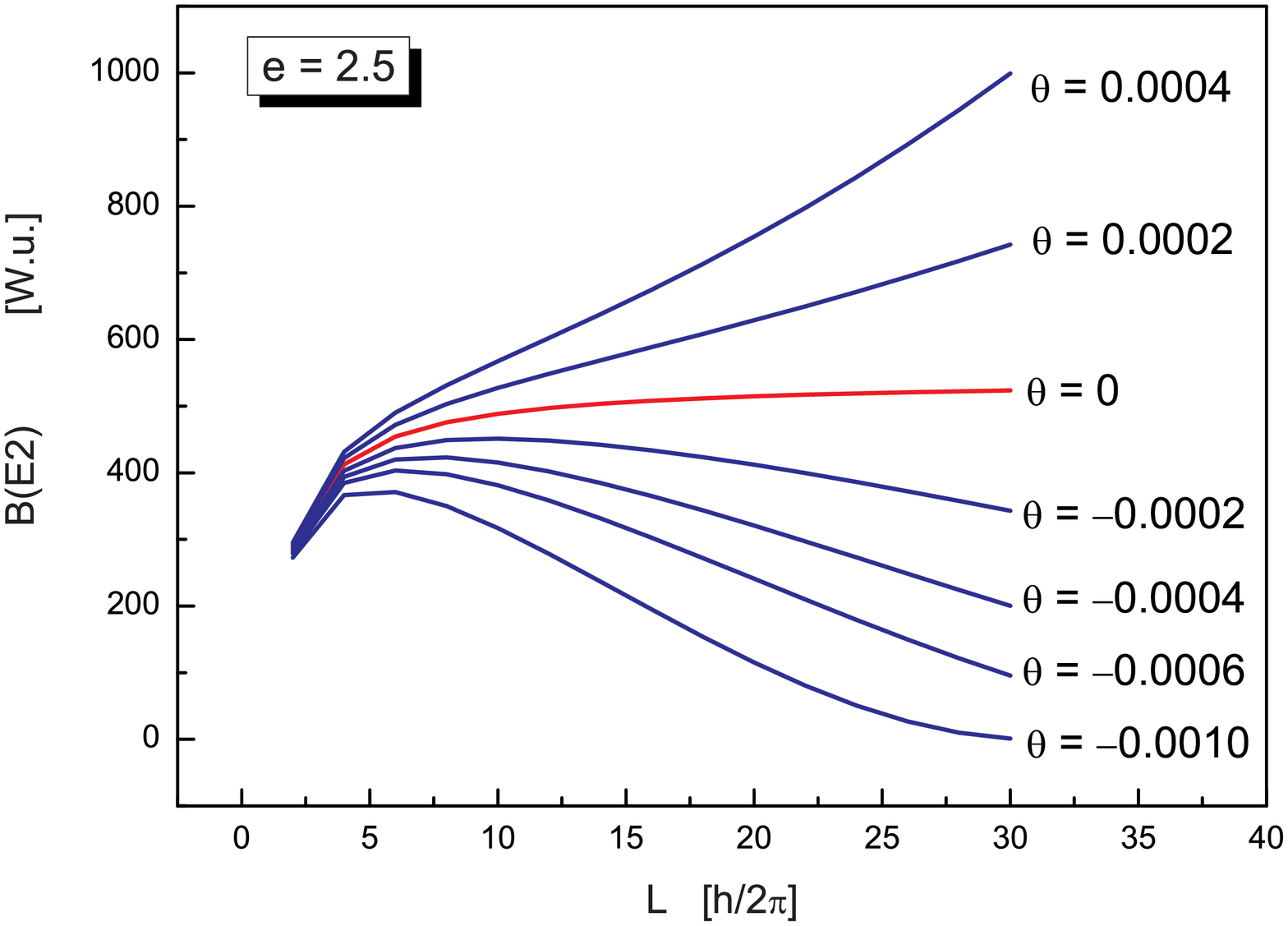}}\hspace{0.5cm}
\centerline{\epsfxsize=3.5in\epsfbox{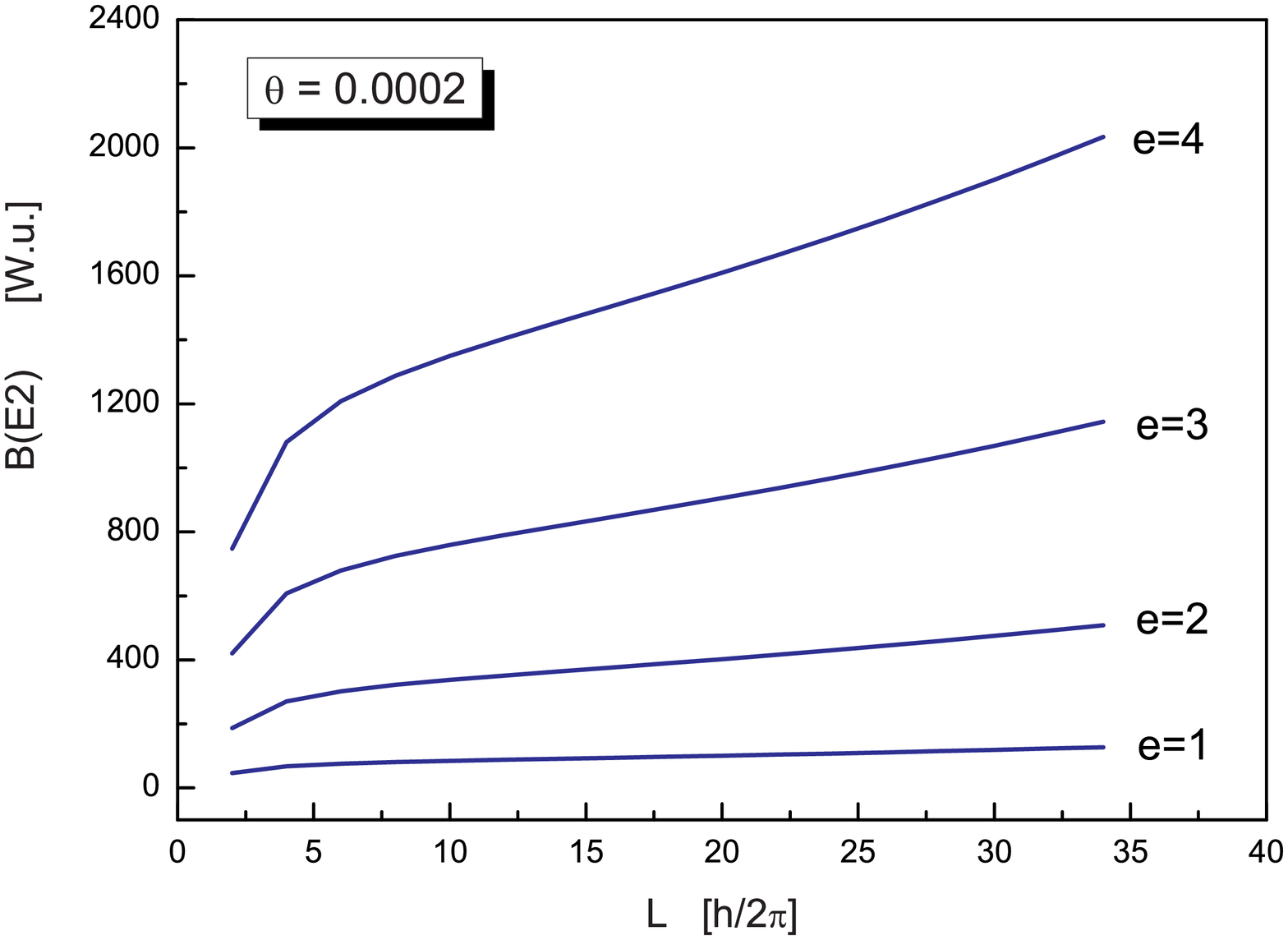}}
\caption{(Color
online) Study of particular dependence of the yrast $B(E2)$
values, using the $E2$ operator (\protect\ref{te2}), as a function
of the parameters $\theta $ and $e$.} \label{varpar}
\end{figure}

Introducing in (\ref{RMEFF}) the above results and the value of the
coefficient $C_{\ \ L\ \quad \ 2\ \quad \ \ \ L+2}^{(0 ,\mu )\ \
(2,0)\ \ \ (0,\mu+2)}$ from \cite{Ver} (the corresponding /fully
stretched \cite{Recoupling}/ $U(6)$ IF for maximal coupling
representations is equal to 1), we finally derive for the $SO(3)$
reduced matrix element
\begin{widetext}
\begin{equation}
\begin{tabular}{l}
$\langle \lbrack N+4],(0,\mu +2);0L+2;00||[F\times
F]_{(0,2)[0]_{2}\quad
00}^{\quad \lbrack 4]_{6}\quad \ 20}||[N],(0,\mu );0L;00\rangle $ \\
\\
$=\left[ \dfrac{(\mu +L+3)(\mu +L+5)(L+1)(L+2)}{(\mu +1)(\mu
+2)(2L+3)(2L+5)}\right]^{1/2}
\sqrt{(N+1)(N+2)(N+3)(N+4)},$\\
\\
$=\sqrt{(2L+1)(2L+2)(2L+3)(2L+4)}$.
\end{tabular}
\label{ME of FF}
\end{equation}
\end{widetext}
where $N=2\mu + \lambda$ and for the last row the yrast condition
$\mu=L$ is taken into account. For the calculation of the matrix
element of $[G\times G]_{(2,0)[0]_{2}\quad 00}^{\quad \lbrack
-4]_{6}\quad \quad 20}$ we use the conjugation
property
\begin{widetext}
\begin{equation}
\begin{tabular}{l}
$\langle \lbrack N-4],(0,\mu -2);0L-20;00|[G\times
G]_{(2,0)[0]_{2}\quad
00}^{\quad \lbrack -4]_{6}\quad \quad 20}|[N],(0,\mu );0L0;00\rangle $ \\
\\
$=(\langle \lbrack N],(0,\mu );0L0;00|[F\times F]_{(0,2)[0]_{2}\quad
00}^{\quad \lbrack 4]_{6}\quad \quad 20}|[N-4],(0,\mu
-2);0L-20;00\rangle
)^{\ast }$ \\
\\
$=C_{(0,\mu -2)[0]_{2}\quad (0,2)[0]_{2}\quad (0,\mu
)[0]_{2}}^{[N-4]_{6\ }\ \ \ \ \ \ \ \ \ \ [4]_{6}\ \ \ \ \ \ \
[N]_{6\ }}C_{\ \ L-2\ \quad \ 2\ \quad \ \ \ L}^{(0 ,\mu-2 )\ \
(0,2)\ \ \ (0,\mu)} C_{L-2,0\ \ 20}^{L,0}
 \sqrt{(N-3)(N-2)(N-1)N}$ \\
 \\ = $ C_{L-2,0\ \
20}^{L,0}\sqrt{(2L-3)(2L-2)(2L-1)2L}. $
\end{tabular}
\label{ME of GG}
\end{equation}%
\end{widetext}
With the help of the above analytic expressions ((\ref{RME of A}),
(\ref{ME of FF}) and (\ref{ME of GG})) for the matrix elements of
the tensor operators forming the $E2$ transition operator we can
calculate the transition probabilities (\ref{deftrpr}) between the
states in the ground  band as attributed to the $SU(3)$
symmetry-adapted basis states of the model (\ref{bast}). It is
obvious that the second term in $T^{E2}$ (\ref{te2}) comes from
the symplectic extension of the model. The behavior of each term
of the transition operator is plotted as a function of the angular
momentum $L$ in Figure \ref{trbe} where for comparison typical
experimental data for the GSB of $^{236}$U are also shown. It can
also be seen that because of the yrast conditions ($\mu =L$), the
well-known parabolic behavior corresponding to the Elliott' s
quadrupole operator is modified and looks like a rigid rotor curve
(see also the curve corresponding to $\theta =0$ in Figure
\ref{varpar}). In this case, the rigid rotor predictions are
asymptotically determined by the ordinary $SO(3)$ Clebsch-Gordan
coefficient. Such type of curve is obtained in the limit of
large-dimensional irreducible representations $2\lambda +
\mu\rightarrow \infty$ when $su(3)$ algebra contracts to the rigid
rotor algebra $rot(3)=[R^{5}]so(3)$ \cite{RoLeRep}. It is obvious
that the experimental points are well reproduced by the modified
$SU(3)$ term up to $L\approx 20$, while for the description of the
states with $L>20$ the symplectic term is appropriate.

In order to see what type of $B(E2)$ behavior can be obtained in
our theoretical predictions we give in Figure \ref{varpar} the
results for various values of the parameters $\theta$ and $e$.
It is clearly seen that the two main  types of $B(E2)$ behavior --
the enhancement or the reduction of the $B(E2)$ values, can be
reproduced. The strongly enhanced values which are an indication
for increased collectivity in the high angular momentum domain are
easily obtained for positive values of the parameter $\theta$. For
negative values of the parameter $\theta$ we obtain behavior
similar to that of the standard $SU(3)$ one and it can be used to
reproduce the well known cutoff effect. Such saturation effect is
also characteristic feature of the IBM based calculations in its
$SU(3)$ limit. Although the coefficient in front of symplectic
term is about four orders of magnitude smaller than the $SU(3)$
contribution to the transition operator its role in reproducing
the correct behavior (with or without cutoff) of the transition
probabilities between the states of the GSB band is very
important.

\section{Application to real nuclei \label{s5}}

In order to prove the correct predictions following from our
theoretical results we apply the theory to real nuclei for which
there is available experimental data for the transition
probabilities \cite{Th232},\cite{U236},\cite{Dy156} between the
states of the ground bands up to very high angular momenta. The
application actually consists of fitting the two parameters of the
transition operator $T^{E2}$ (\ref{te2}) to the experiment for each
of the considered bands.

\begin{figure}[h]
\centerline{\epsfxsize=3.5in\epsfbox{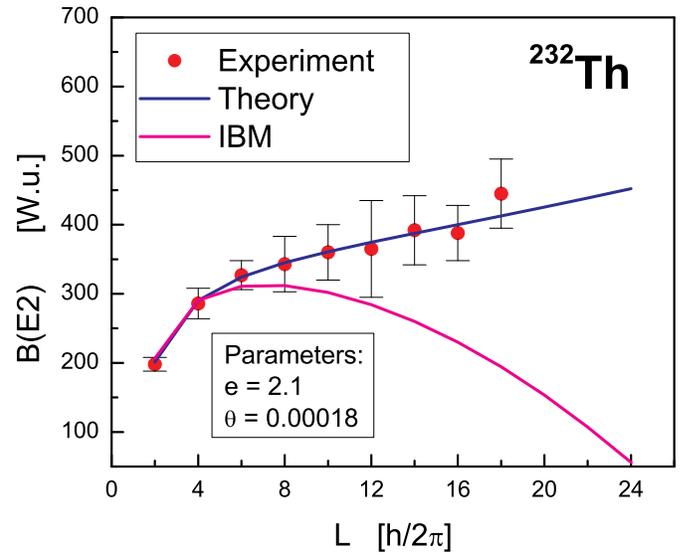}}
%\centering
%\includegraphics[height=8cm]{Th232.eps}
\caption{(Color online) Comparison of theoretical and experimental
values for the $B(E2)$ transition probabilities for the $^{232}Th$.}
\label{Th}
\end{figure}

As a first example we consider the intraband $B(E2)$ transitions
in the GSB for the nucleus $^{232}Th$. The experimental data for
it are compared with the corresponding theoretical results of the
Symplectic IVBM and the $SU(3)$ limit of the IBM in the Figure
\ref{Th}. We see the standard $SU(3)$ behavior for the latter and
since the IBM involves small number of quadrupole bosons the
cutoff effect is observed at low spins and hence only the
transitions between the first few excited states are well
reproduced by it. From Figure \ref{Th} one can see the enhanced
$B(E2)$ values in the high spin region and the good reproduction
of the experimental data \cite{Th232} by our theoretical
predictions.

\begin{figure}[h]
\centerline{\epsfxsize=3.5in\epsfbox{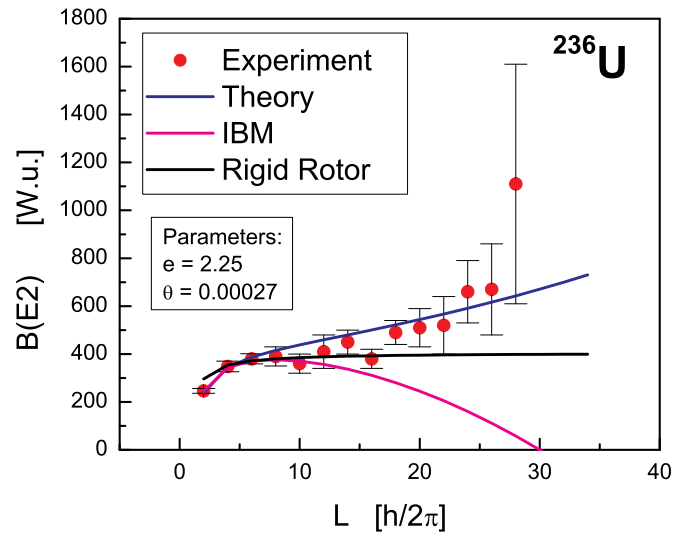}}
% \centering
%\includegraphics[height=8cm]{U236.eps}
\caption{(Color online) The same as in the Fig. \protect\ref{Th} but
for $^{236}U$ .} \label{Ur}
\end{figure}

Next the $^{236}U$ case is presented. For it there are a lot of
experimental data, reaching high angular momenta up to $L=28$
\cite{U236}. The $B(E2)$ values for transitions between members of
the GSB compared with the theoretical results of the IVBM, the IBM
and the rigid rotor are shown in Figure \ref{Ur}. One can see that
the IBM works well for the transitions between the first excited
states ($L=2-10$). The rigid rotor describes well the experimental
states in the middle spin region ($L=4-16$), while for the high
spins the $B(E2)$ values must be enhanced due to the observed
collectivity excess. Thus, at high spins in the yrast band the
calculations of the IBM and the rigid rotor model cannot reproduce
the fine structure of the $B(E2)$ data. As was mentioned in the
preceding section, such an enhancement can be obtained for slightly
positive values of the parameter $\theta$ in the transition operator
$T^{E2}$ (\protect\ref{te2}) (see Figure \ref{varpar}). From Figure
\ref{Ur} one can see that the experimental points, lay very close to
the theoretical curves.

\begin{figure}[h]
\centerline{\epsfxsize=3.5in\epsfbox{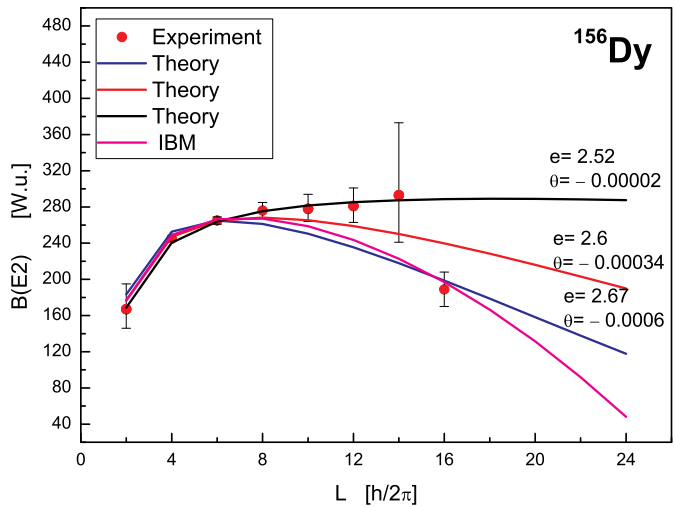}}
% \centering
%\includegraphics[height=8cm]{Dy156com.eps}
\caption{(Color online) Comparison of theoretical and experimental
values for the $B(E2)$ transition probabilities for several closed
values of the parameters $\theta$ and $e$ for the nucleus
$^{156}Dy$.} \label{Dyb}
\end{figure}

As a final example we consider the $^{156}Dy$ nucleus. The results
(values of $e$ and $\theta $) obtained for the yrast band compared
with that of the IBM and the experimental data \cite{Dy156} are
presented on Figure \ref{Dyb}. From it the saturation effect is
clearly observed at $L=16$. The calculated results (for the IVBM
with negative value of the parameter $\theta $) illustrate
characteristics of the generic $B(E2)$ curve discussed in connection
with Figure \ref{varpar}. We see that the two models (for IVBM$-$
the blue curve) give identical results with about the same level of
accuracy. As one can see the better overall reproduction of the
experimental data can be obtained if the parameters $\theta $ and
$e$ are slightly modified which is also illustrated in the Figure
\ref{Dyb}.

From the presented examples we see how sensitive is the theory to
the term coming from the symplectic extension and in particular
from the sign of the parameter $\theta $ entering in the
transition operator (\protect\ref{te2}).

\section{Conclusions \label{s6}}

In the present paper we investigate the tensor properties of the
algebra generators of $Sp(12,R)$ with respect to the reduction chain
(\ref{RotLimit}). $Sp(12,R)$ is the group of dynamical symmetry of
the IVBM and the considered chain of subgroups was applied in
\cite{Sp12U6} for the description of positive and negative parity
bands in well deformed nuclei. The basis states of the model
Hamiltonian are also classified by the quantum numbers corresponding
to the irreducible representations of the subgroups from the chain
and in this way is constructed the symmetry adapted basis in this
limit of the model. The action of the symplectic generators as
transition operators between the basis states is analyzed.
Analytical expressions for the matrix elements of $Sp(12,R)$
generators in the $U(6)$ symmetry-adapted basis are obtained as
well.

In the present new application of the rotational limit of the
symplectic extension of the IVBM, the model was tested on the more
complicated and complex problem of reproducing the $B(E2)$
transition probabilities between the states of the ground  band up
to very high spins. In developing the theory the advantages of the
algebraic approach were used first for the proper assignment of the
basis states  to the experimentally observed states of the
collective bands. Here the construction of the $E2$ transition
operator as linear combination of tensor operators representing the
generators of the subgroups of the respective chain is a basic
result that allows the application of a specific version of the
Wigner - Eckart theorem and consecutively leads to analytic results
for their matrix elements in $U(6)$ symmetry-adapted basis, that
give the transition probabilities.

Analyzing the terms taking part in the construction of the $E2$
transition operator the important role of the symplectic extension
of the model is revealed. In the application to real nuclei the
parameters of the transition operator are evaluated in a fitting
procedure for GSB of the considered nuclei. The experimental data
for the presented examples is reproduced rather well, although the
results are very sensitive to the values of the parameters.

A further investigation of these bands in other nuclei from other
nuclear regions will clarify better the development of
collectivity in the symplectic extension of the IVBM, for which
more experiment on transition probabilities is needed. The
presented approach is rather general and universal and can be used
for the calculation of transitions in other collective bands, in
particular in the similarly constructed negative parity bands and
the excited $\beta -$ bands, which are of great interest lately in
the nuclear structure.

\section*{Acknowledgments}

The authors acknowledge the useful discussions of this work with
Dr. V. P. Garistov. This work was supported by the Bulgarian
National Foundation for scientific research under Grant Number
$\Phi -1501$.

\end{document}